\documentclass[fleqn,usenatbib,useAMS]{mnras}

\usepackage{graphicx}	
\usepackage{amsmath}	
\usepackage{amssymb}	

\usepackage{multicol}        
\usepackage{bm}		
\usepackage{pdflscape}	
\usepackage{epsf}
\usepackage{epstopdf}
\usepackage{dcolumn}   
\usepackage{color}
\usepackage{multirow}   
\usepackage{hyperref}
\usepackage{widetext}
\usepackage[dvipsnames]{xcolor}
\hypersetup{colorlinks=true,citecolor=blue, linkcolor=red, urlcolor=purple}

\DeclareRobustCommand{\VAN}[3]{#2}
\let\VANthebibliography\thebibliography
\def\thebibliography{\DeclareRobustCommand{\VAN}[3]{##3}\VANthebibliography}

\usepackage[T1]{fontenc}
\usepackage{ae,aecompl}

\usepackage{newtxtext,newtxmath}

\title[Spatial Curvature and Thermodynamics]{Spatial Curvature and Thermodynamics}

\author[N. Banerjee, P. Mukherjee, D. Pav\'{o}n]{
	Narayan Banerjee$^{1}$ \thanks{Contact e-mail: \href{mailto:narayan@iiserkol.ac.in}{narayan@iiserkol.ac.in}}, 
	Purba Mukherjee$^{2}$ \thanks{Contact e-mail: \href{mailto:purba16@gmail.com}{purba16@gmail.com}}
	and 
	Diego Pav\'{o}n$^{3}$ \thanks{Contact e-mail: \href{mailto:diego.pavon@uab.es}{diego.pavon@uab.es}}
	\\
	\vspace{-0.2cm}
	\\
$^{1}$Department of Physical Sciences, Indian Institute of Science Education and Research, Mohanpur, West Bengal - 741246 India\\
$^{2}$Physics and Applied Mathematics Unit, Indian Statistical Institute, Kolkata - 700108, India\\
$^{3}$Departamento de F\'{\i}sica, Facultad de Ciencias, Universidad Aut\'{o}noma de Barcelona, 08193 Bellaterra	(Barcelona), Spain
}

\date{Accepted 2023 March 22. Received 2023 March 18; in original form 2023 January 23}

\pubyear{????}

\begin{document}
\label{firstpage}
\pagerange{\pageref{firstpage}--\pageref{lastpage}}
\maketitle

\begin{abstract}
Reasonable parametrizations of the current Hubble data set of the
expansion rate of our homogeneous and isotropic universe, after
suitable smoothing of these data, strongly suggest that the area
of the apparent horizon increases irrespective of whether the
spatial curvature of the metric is open, flat or closed. Put in
another way, any sign of the spatial curvature appears consistent
with the second law of thermodynamics.
\end{abstract}


\section{Introduction \label{sec:1}}
\noindent Homogeneous and isotropic cosmological models are most
aptly described by the Friedmann-Lema\^{i}tre-Robertson-Walker
(FLRW) space-time metric,
\begin{equation}
ds^{2} = - dt^{2}\, + a^{2}(t)\, \left[\frac{dr^{2}}{1-kr^{2}} \,
+ \, r^{2}\left(d \theta^{2} \, + \, \sin^{2} \theta  d \phi^{2}
\right) \right] \, .
\label{metric}
\end{equation}
The spatial curvature index, $k \in \{-1, 0, 1\}$, indicates whether
the spatial part of the metric is open (negatively curved, i.e., hyperbolic),
flat, or closed (positively curved).\\

\noindent This constant index, like the scale factor $a(t)$, is
not a directly observable quantity. In principle, however, it can
be determined through the knowledge of the spatial curvature
density parameter, $ \Omega_{k} \equiv -k/(a^{2}H^{2})$, which is
accessible to observation, albeit indirectly. Recent estimates of
the latter, assuming the universe correctly described by the
$\Lambda$CDM model only suggests that its present absolute value
is small ($\mid \Omega_{k0}\mid \sim 10^{-2}$ or less
\citep{komatsu2011, planck2013, aghanim2018, vagnozzi2021a,
vagnozzi2021b, dawhan2021, ratra2018, handley2021,
narayan2022, bel2022, akarsu2023}). More recent measurements, not
based in the aforesaid model, hint that the universe may not be
flat (i.e., $k \neq 0$) \citep{ratra2018, handley2021, narayan2022,
bel2022}. So,  the sign of $k$ is rather uncertain. \\

\noindent Based on the history of the Hubble factor $H(z)$, (where
$H = \dot{a}/a$ ) it has been recently argued \citep{mnras484} that the
universe is a thermodynamic system in the sense  that it satisfies the second
law of thermodynamics (the first law is guaranteed by Einstein
equations while the third law may not apply). As we shall see
soon, the possibilities $k = +1$ and $k = 0$ are consistent with
the second law while this is not so obvious for the third
possibility; in principle $k = -1$ may or may not be compatible
with the second law. The aim of our study is to resolve this
uncertainty. To this end we shall resort to the 60 $H(z)$ data
currently available alongside their $1 \sigma$ confidence
intervals, the parametrized graph of this history after smoothing
the data set by a Gaussian Process \citep{rasmussen2006}, the FLRW
metric and the expression of the area of the apparent horizon (Eq.
(\ref{area}) below).  As it turns out, the  $k = -1$ possibility
(open spatial sections) appears also compatible with the second
law of thermodynamics. The paper is organized as follows: In
section \ref{sec:2}  we write the second law at cosmic scales in terms of
the deceleration parameter and $\Omega_{k}$. In section \ref{sec:3} we
present the $H(z)$ data set and the smoothing process. In section \ref{sec:4}
we introduce four parametrizations of the Hubble factor and study
whether they are compatible with the second law. In section \ref{sec:5} we
consider the impact of the Hubble constant priors on our results.
Finally, in section \ref{sec:6} we summarize our findings. \\

\noindent  As usual, a subindex zero attached at any given
quantity indicates that the quantity is to be evaluated at the present epoch.
Likewise, we recall, for future convenience,  that after fixing $a_{0} =1$, the
redshift, $z$,  of any given luminous source is related to the
corresponding scale factor by $\, 1+z = a^{-1}$.  In our units $\,
c = G = \hbar = 1$.

\section{The second law} \label{sec:2}

\begin{table}
	\centering
	\caption{Recent Hubble parameter compilation from cosmic chronometers.}
	\label{tab:Hubble_CC}
	\renewcommand{\tabcolsep}{0.4pc} 
	\renewcommand{\arraystretch}{1.1} 
	\begin{tabular}{ c c c }
		\hline 
		$z$ & $H(z)$ [km Mpc$^{-1}$ s$^{-1}$] & Ref. \\
		\hline 
		0.09    &  ~69 $\pm$  12   & \multirow{11}*{\citet{stern2010}} \\
		0.17   &  ~83 $\pm$  8~ & \\
		0.27   &  ~77 $\pm$  14   & \\
		0.4 &  ~95 $\pm$  17   & \\
		0.48   &  ~97 $\pm$  62   & \\
		0.88   &  ~90 $\pm$  40   & \\
		0.9 &  117 $\pm$  23   & \\
		1.3 &  168 $\pm$  17   & \\
		1.43   &  177 $\pm$  18   & \\
		1.53   &  140 $\pm$  14   & \\
		1.75   &  202 $\pm$  40   & \\
		\hline
		0.1797 & ~75 $\pm$ 4~   & \multirow{8}*{\citet{moresco2012}} \\
		0.1993 & ~75 $\pm$ 5~   & \\
		0.3519 & ~83 $\pm$ 14   & \\
		0.5929 & 104 $\pm$ 13   & \\
		0.6797 & ~92 $\pm$ 8~   & \\
		0.7812 & 105 $\pm$ 12   & \\
		0.8754 & 125 $\pm$ 17   & \\
		1.037  & 154 $\pm$ 20   & \\
		\hline
		0.07   &  69.0  $\pm$  19.6 & \multirow{4}*{\citet{zhang2014}} \\
		0.12   &  68.6  $\pm$  26.2 & \\
		0.2 &  72.9  $\pm$  29.6 & \\
		0.28   &  88.8  $\pm$  36.6 & \\
		\hline
		1.363  & 160.0 $\pm$ 33.6  & \multirow{2}*{\citet{moresco2015b}} \\
		1.965  & 186.5 $\pm$ 50.4  & \\
		\hline
		0.3802 & ~83.0 $\pm$ 13.5  & \multirow{5}*{\citet{moresco2016}} \\
		0.4004 & ~77.0 $\pm$ 10.2  & \\
		0.4247 & ~87.1 $\pm$ 11.2  & \\
		0.4497 & ~92.8 $\pm$ 12.9  & \\
		0.4783 & ~80.9 $\pm$ 9~~~ & \\
		\hline
		0.47   & ~~89 $\pm$ 49.6    & \citet{ratsim2017} \\
		\hline
		0.75 & ~~98.8 $\pm$ 33.6 & \citet{borghi2022}\\
		\hline 
	\end{tabular}
\end{table}

\noindent Given the strong connection between gravity and
thermodynamics \citep{jakob1974, jakob1975, steven, jacobson, paddy}, it is natural
to expect that the universe behaves as a normal thermodynamic
system \citep{mnras484} whereby it must tend to a state of
maximum entropy in the long run \citep{grg1, grg2}. \\

\noindent A basic standpoint is that the entropy of the universe
is dominated by entropy of the cosmic horizon. In fact its
entropy ($\sim 10^{132}\rm{k_{B}}$) exceeds by far the combined entropies
of super-massive black holes, the cosmic
microwave background radiation, the neutrino sea, etc \citep{apj-Egan}. As
cosmic horizon, we take the apparent horizon rather than other
possible choices, since the laws of thermodynamics are fulfilled on
it \citep{bye}. Its entropy (proportional to its area) is given by
$ \, S_{\cal A} = \rm{k_{B}} \pi \, \tilde{r}^{2}_{\cal
	A}/{\ell_{p}}^2 \;$ \citep{bak-rey, cai2008}, with $\tilde{r}_{\cal
	A} = (H^{2}+ka^{-2})^{-1/2} \,$ as the radius of the horizon. \\

\noindent Obviously, the area of the apparent horizon,
\begin{equation}
{\cal A} = \frac{4 \pi}{H^{2} \, + \, \frac{k}{a^{2}}} \, ,
\label{area}
\end{equation}
depends on the Hubble factor and  spatial curvature.
\\  \

\noindent By the second law of thermodynamics $S_{\cal A}' \geq 0$, thus
\begin{equation}
{\cal A}' = - \frac{{\cal A}^{2}}{8 \pi^2} \, \left(H H' \, - \,
\frac{k}{a^{3}}\right) \geq 0 \quad \Rightarrow  \quad H H' \leq
\frac{k}{a^{3}}\, , \label{eq:Aprime}
\end{equation}
where the prime indicates $d/da$. \\

\noindent The above inequality tells us: $(i)$ if $H'$ is or has
been positive at any stage of cosmic expansion (excluding,
possibly, the pre-Planckian era), then $k = +1$, $(ii)$  $H' <0$
may, in principle, be compatible with any sign of $k$, and $(iii)$
the possibilities $k = +1$  and  $k = 0$ are consistent with the
second law. In particular the data analysis carried out by the 2018 
Planck mission produced the 99$\%$ probability region 
$ -0.095 < \Omega_{k0} < -0.007$ on the spatial curvature parameter 
\citep{aghanim2018}. Note that this bound on $\Omega_{k0}$ is fully 
compatible with the second law because it corresponds 
to the possibility $k = +1$. However, the third possibility, $k = -1$, 
may or may not be compatible.\\

\begin{table}
	\centering
	\caption{Recent Hubble parameter compilation from the radial BAO \& galaxy clustering.}
	\label{tab:Hubble_BAO}
	\renewcommand{\tabcolsep}{0.3pc} 
	\renewcommand{\arraystretch}{1.1} 
	\begin{tabular}{ c c c c }
		\hline 
		$z$ & $H(z)$ [km Mpc$^{-1}$ s$^{-1}$] & ${r_d}^{\text{fid}}$ [Mpc] & Ref. \\
		\hline 
		0.24   & 79.69 $\pm$ 2.99 & \multirow{3}*{153.3} &\multirow{3}*{\citet{gaztanaga2009}} \\
		0.34   & 83.8 $\pm$ 3.66 &  & \\
		0.43   & 86.45 $\pm$ 3.97 &  & \\
		\hline
		0.44   & 82.6 $\pm$ 7.8  & \multirow{3}*{$-$} &\multirow{3}*{\citet{blake2012}} \\
		0.6 & 87.9 $\pm$ 6.1  & &\\
		0.73   & 97.3 $\pm$ 7 & &\\
		\hline
		0.3 & 81.7 $\pm$ 6.22 & 152.76 & \citet{oka2013} \\
		\hline
		0.35   & 82.7 $\pm$ 9.1  & 152.76 & \citet{chuang2013} \\
		\hline
		0.57   & 96.8 $\pm$ 3.4  & 147.28 & \citet{anderson2014} \\
		\hline
		0.31   & 78.17 $\pm$ 4.74 & \multirow{9}*{147.74} & \multirow{9}*{\citet{wang2016}} \\
		0.36   & 79.93 $\pm$ 3.39 & &\\
		0.40   & 82.04 $\pm$ 2.03 & &\\
		0.44   & 84.81 $\pm$ 1.83 & &\\
		0.48   & 87.79 $\pm$ 2.03 & &\\
		0.52   & 94.35 $\pm$ 2.65 & &\\
		0.56   & 93.33 $\pm$ 2.32 & &\\
		0.59   & 98.48 $\pm$ 3.19 & &\\
		0.64   & 98.82 $\pm$ 2.99 & &\\
		\hline
		0.38   & 81.5 $\pm$ 1.9  & \multirow{3}*{147.78} &\multirow{3}*{\citet{alam2016}} \\
		0.51   & 90.4 $\pm$ 1.9  &  &\\
		0.61   & 97.3 $\pm$ 2.1  &  &\\
		\hline
		0.978   & 113.72 $\pm$ 14.63  & \multirow{4}*{147.78} &\multirow{4}*{\citet{zhao2019}} \\
		1.23   & 131.44 $\pm$ 12.42  &  & \\
		1.526   & 148.11 $\pm$ 12.71  &  & \\
		1.944   & 172.63 $\pm$ 14.79  &  & \\
		\hline
		2.33   & 224 $\pm$ 8 &  147.33 & \citet{bautista2017} \\
		\hline
		2.34   & 222 $\pm$ 7    &  147.7 & \citet{delubac2014} \\
		\hline
		2.36   & 226 $\pm$ 8  & 147.49 & \citet{font-ribera2014} \\
		\hline 
	\end{tabular}
\end{table}

\noindent The inequality $H H' \leq \frac{k}{a^{3}} $ can be recast as
\begin{equation}
1 + q \geq \Omega_k \, ,
\label{q_inequality}
\end{equation}
where $q = -\ddot{a}/(a H^{2}) = -(1 + \dot{H} H^{-2})$ is the
deceleration parameter. \\

\noindent Equation (\ref{q_inequality}) expresses the second law
of thermodynamics at cosmic scales \citep{mnras484}.
\\  

\begin{figure*}
	\centering
	\includegraphics[width=0.85\textwidth]{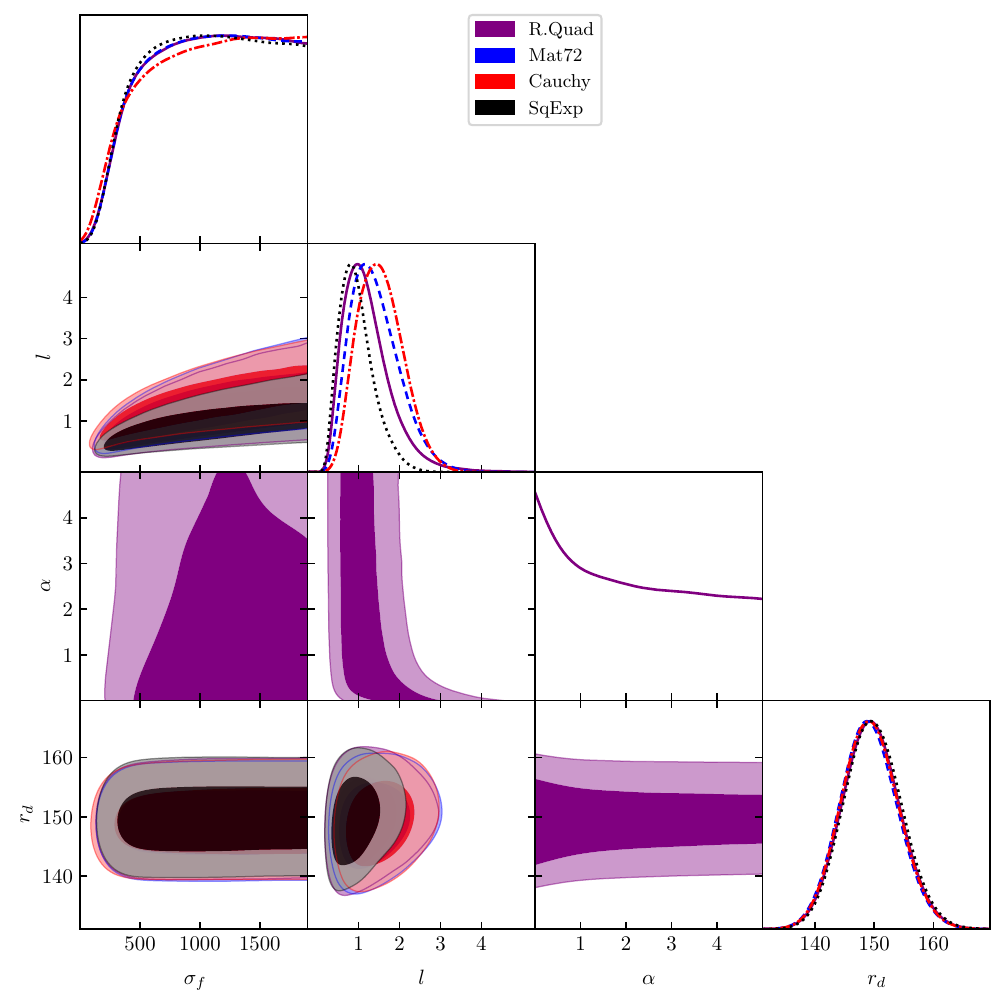}
	\caption{Contour plots for GP hyperparameter space and the comoving sound horizon at drag epoch whose radius, $r_d$,
		is measured in megaparsecs.}
	\label{hyper_plot}
\end{figure*}

\noindent To see that it is perfectly  reasonable, let us assume the universe dominated by a
perfect fluid of energy density $\rho$ and hydrostatic pressure
$P$. The corresponding Einstein equations can be written as,
\begin{equation}
H^{2} + \frac{k}{a^{2}} = \frac{8 \pi}{3} \rho   \qquad {\rm and} \qquad
\frac{\ddot{a}}{a} = - \frac{4 \pi}{3}(\rho + 3 P).
\label{eq:einstein1&2}
\end{equation}
If the fluid is dust ($P = 0$), one has $\, q > 0$,
and these equations lead to $\, \Omega_{k} = 1 - 2q$.
\\  \

\noindent If the second law holds, i.e., $1 + q \geq \Omega_{k}$,
it follows that  $\, q \geq 0$ which is consistent with the fact that
a pressureless matter dominated universe decelerates ($ q > 0$).
\\  \

\noindent By contrast, should the second law  not hold, $1 + q <
\Omega_{k}$, the absurd result $0<-q$ (namely, zero less than a
negative quantity), would follow.
\\  \

\noindent Likewise, if the equation of state of the fluid filling
the universe were $P = w \rho$ with $\, w \geq -1$ the expression
$1 + q \geq \Omega_{k}$ also proves to be consistent with the
corresponding sign  of the deceleration parameter and not so, if $1 +
q < \Omega_{k}$. Values of $w$ lower that $-1$ correspond to
phantom fields. As is well known these fields are unstable both
classically \citep{dabrowski2015} and quantum
mechanically\footnote{Should $H'$ be positive, the FLRW universe
	could neither be flat nor open, just closed ($k = +1$) which
	would be puzzling in view of the current data that clearly allow
	for $k = 0$.} {\,} \citep{cline2004}.
\\  \

\noindent Further,  by combining the Friedmann equation in the case of a
universe dominated by pressureless matter, the cosmological
constant and spatial curvature with the acceleration equation
(\ref{eq:einstein1&2}.b), we obtain
\begin{equation}
\frac{\ddot{a}}{a} \, - \, \left( \frac{\dot{a}}{a}\right)^{2} \,
- \frac{k}{a^{2}} = -4 \pi  \rho_{m}\, .
\label{eq:acceleration}
\end{equation}
The latter can be rewritten as
\begin{equation}
1 + q = \Omega_{k} + \textstyle{3\over{2}} \Omega_{m} \,  \qquad
\qquad (\Omega_{m} = 8 \pi \rho_{m}/(3 H^{2})),
\label{eq:twoomegas}
\end{equation}
which makes apparent the reasonableness of the expression $\, 1 + q \geq
\Omega_{k}$. Moreover, the above shows the compatibility of
the said expression with general relativity.



\begin{figure*}
	\begin{minipage}{0.325\textwidth}
		\begin{minipage}{\textwidth}
			\includegraphics[width=\textwidth]{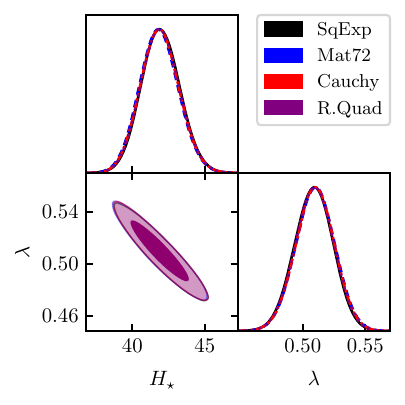}\\ \vspace{-0.8cm}
			\begin{center}
				(a) M1
			\end{center}
			\label{fig:contour_M1}
		\end{minipage}
		\begin{minipage}{\textwidth}
			\includegraphics[width=\textwidth]{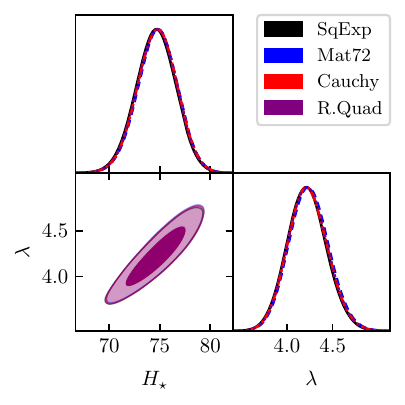}\\ \vspace{-0.8cm}
			\begin{center}
			(c) M3
			\end{center}
			\label{fig:contour_M3}
		\end{minipage}
	\end{minipage}
	\begin{minipage}{0.625\textwidth}
		\includegraphics[width=\textwidth]{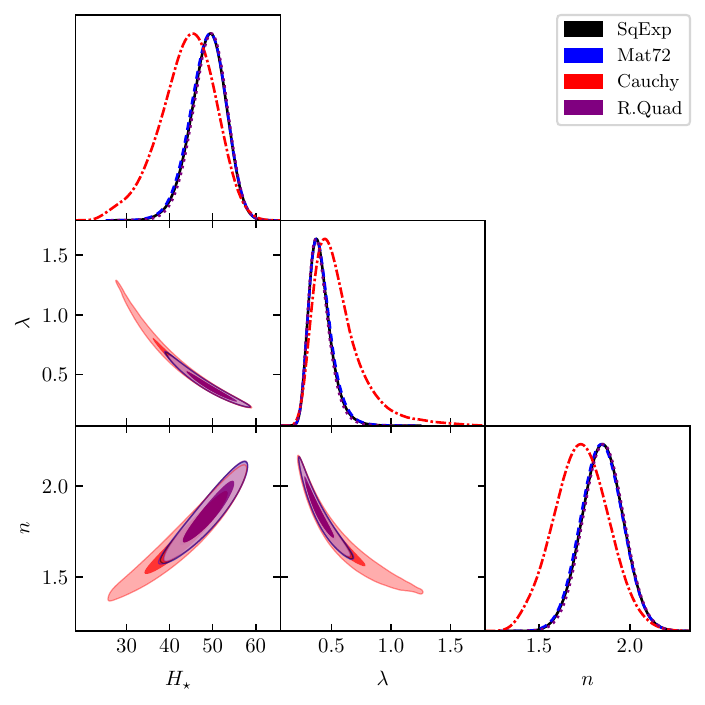}\\ \vspace{-0.8cm}
			\begin{center}
			(b) M2
			\end{center}
		\label{fig:contour_M2}
	\end{minipage}%
	\hfill
	\caption{Contour plots for (a) first $H(a)$ model parameters, (b) second $H(a)$ model parameters, and (c) third $H(a)$ model parameters,
		using different choices of the GP kernel, where $H_\star$ is in units of km Mpc$^{-1}$ s$^{-1}$.} \label{fig:contour}
\end{figure*}


\section{Hubble data set and smoothing process}\label{sec:3}
\noindent Our data set consists of: $(i)$ recent 32 cosmic
chronometer (CC) $H(z)$ measurements, presented in Table
\ref{tab:Hubble_CC}, which do not assume any particular
cosmological model, and $(ii)$ the 28 $H(z)$ measurements from
baryon acoustic oscillations (BAO) from different galaxy surveys,
listed in Table \ref{tab:Hubble_BAO}. In both cases, the datasets are given
with their $1\sigma$ confidence interval. Because the BAO measurements are not
entirely model-independent, particularly due to the assumption of
a fiducial radius of the comoving sound horizon,
${r_d}^{\text{fid}}$, we adopted a full marginalization over the
GP hyperparameter space (see below) with the comoving sound
horizon at drag epoch $r_d$ for the BAO data set as free
parameter. This results in a model-independent Hubble data set
from CC and the calibrated BAO.
\\  \


\begin{table*}
	\caption{{\small Parameter values for the first (M1) parametrization corresponding to each kernel, where $H_\star$ and $H_0$ are in units of km Mpc$^{-1}$ s$^{-1}$.}}
	\begin{center}
		\resizebox{\textwidth}{!}{\renewcommand{\arraystretch}{1.15} \setlength{\tabcolsep}{20 pt} \centering
			\begin{tabular}{l c c c c c}
				\hline 
				$k(a, \tilde{a})$ & $H_{\star}$ & $\lambda$ & $H_0$ & $q_0$ & $a_{t}$\\
				\hline
				Sq.Exp & $41.926^{+1.332}_{-1.292}$ & $0.509^{+0.015}_{-0.015}$ & $69.751^{+ 1.229}_{-1.211}$ & $-0.488^{+0.014}_{-0.013}$ & $0.509$ \\
				
				Mat.7/2 & $41.818^{+1.345}_{-1.311}$ & $0.510^{+0.015}_{-0.016}$ & $69.622^{+1.274}_{-1.267}$ & $-0.487^{+0.014}_{-0.014}$ & $0.510$\\
				
				Cauchy & $41.879^{+1.339}_{-1.304}$ & $0.510^{+0.015}_{-0.015}$ & $69.712^{+1.273}_{-1.258}$ & $-0.487^{+0.014}_{-0.014}$ & $0.510$ \\
				
				R.Quad & $ 41.893^{+1.340}_{-1.303}$ & $0.509^{+0.015}_{-0.015}$ & $69.722^{+1.289}_{-1.260}$ & $-0.488^{+0.014}_{-0.014}$ & $0.509$ \\
				
				\hline 
			\end{tabular}
		}
	\end{center}
	\label{tab:M1_table}
\end{table*}

\begin{table*}
	\caption{{\small Parameter values for the second (M2) parametrization, where $H_\star$ and $H_0$ are in units of km Mpc$^{-1}$ s$^{-1}$.}}
	\begin{center}
		\resizebox{\textwidth}{!}{\renewcommand{\arraystretch}{1.15} \setlength{\tabcolsep}{15 pt} \centering
			\begin{tabular}{l c c c c c c}
				\hline 
				$k(a, \tilde{a})$ & $H_{\star}$ & $\lambda$ & $n$ & $H_0$ & $q_0$ & $a_{t}$\\
				\hline
				Sq.Exp & $49.107^{+3.835}_{-4.247}$ & $0.397^{+0.104}_{-0.077}$ & $1.849^{+0.113}_{-0.113}$ & $68.664^{+1.477}_{-1.563}$ & $-0.469^{+0.044}_{-0.047}$ & $0.556$ \\
				
				Mat.7/2 & $48.973^{+3.927}_{-4.361}$ & $0.401^{+0.108}_{-0.079}$  & $1.845^{+0.115}_{-0.114}$  &  $68.626^{+1.538}_{-1.543}$ & $-0.468^{+0.045}_{-0.047}$ & $0.555$\\
				
				Cauchy & $44.486^{+5.934}_{-6.967}$ & $0.513^{+0.232}_{-0.142}$  & $1.731^{+0.152}_{-0.151}$ &  $67.317^{+2.063}_{-2.084}$ & $-0.447^{+0.068}_{-0.073}$ & $0.567$ \\
				
				R.Quad & $49.342^{+3.762}_{-4.121}$ & $0.392^{+0.100}_{-0.075}$ & $1.855^{+0.112}_{-0.111}$  &  $68.692^{+1.494}_{-1.520}$ & $-0.472^{+0.044}_{-0.045}$ & $0.554$ \\
				
				\hline 
			\end{tabular}
		}
	\end{center}
	\label{tab:M2_table}
\end{table*}

\begin{table*}
	\caption{{\small Parameter values for the third (M3) parametrization corresponding to each kernel, where $H_\star$ and $H_0$ are in units of km Mpc$^{-1}$ s$^{-1}$.}}
	\begin{center}
		\resizebox{\textwidth}{!}{\renewcommand{\arraystretch}{1.15} \setlength{\tabcolsep}{20 pt} \centering
			\begin{tabular}{l c c c c c}
				\hline 
				$k(a, \tilde{a})$ & $H_{\star}$ & $\lambda$ & $H_0$ & $q_0$ & $a_{t}$\\
				\hline
				Sq.Exp & $74.635^{+1.938}_{-1.970}$ & $4.215^{+0.216}_{-0.210}$ & $74.750^{+ 1.772}_{-1.770}$ & $-0.842^{+0.020}_{-0.019}$ & $0.546$ \\
				
				Mat.7/2 & $74.798^{+1.928}_{-1.964}$ & $4.237^{+0.220}_{-0.212}$ & $74.917^{+1.791}_{-1.721}$ & $-0.847^{+0.020}_{-0.018}$ & $0.544$\\
				
				Cauchy & $74.751^{+1.931}_{-1.975}$ & $4.224^{+0.221}_{-0.210}$ & $74.878^{+1.806}_{-1.779}$ & $-0.846^{+0.020}_{-0.019}$ & $0.545$ \\
				
				R.Quad & $ 74.735^{+1.922}_{-1.978}$ & $4.225^{+0.218}_{-0.210}$ & $74.862^{+1.763}_{-1.752}$ & $-0.846^{+0.020}_{-0.019}$ & $0.545$ \\
				
				\hline 
			\end{tabular}
		}
	\end{center}
	\label{tab:M3_table}
\end{table*}


\begin{figure*}
	\centering
	\includegraphics[width=0.485\textwidth]{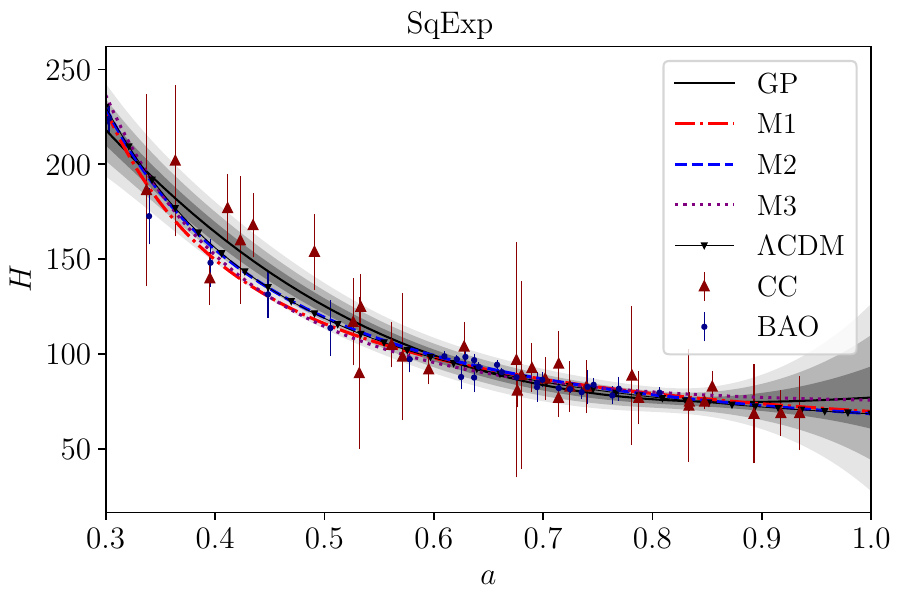}
	\includegraphics[width=0.485\textwidth]{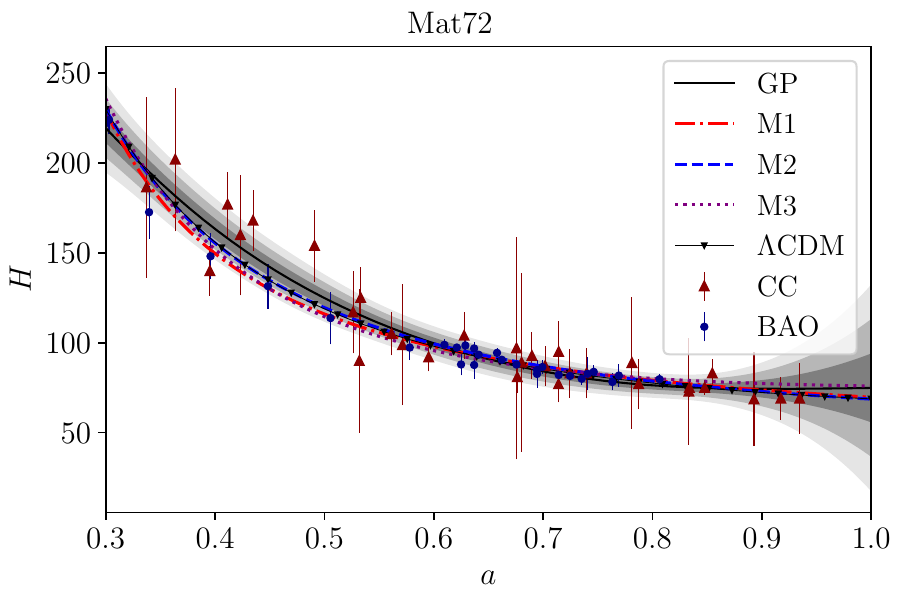}\\
	\includegraphics[width=0.485\textwidth]{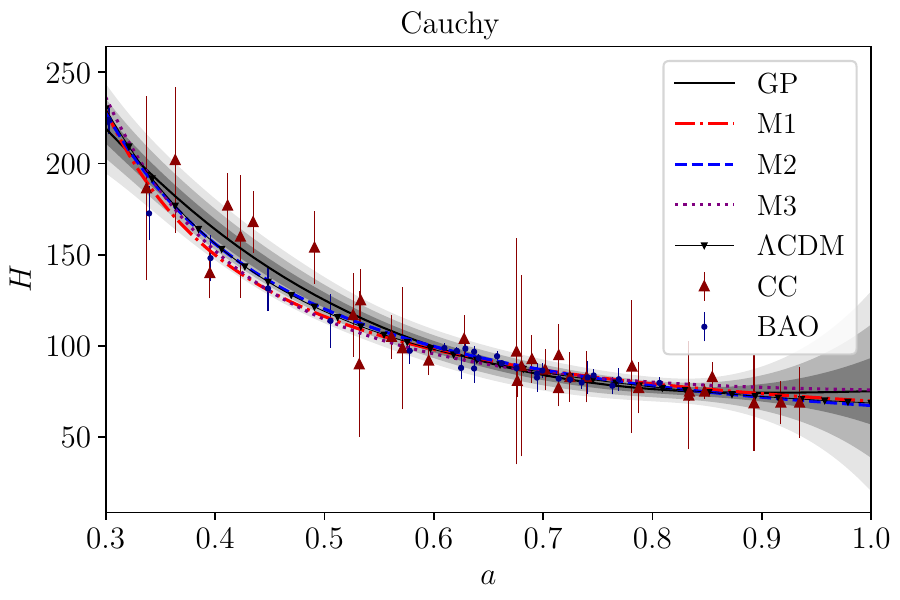}
	\includegraphics[width=0.485\textwidth]{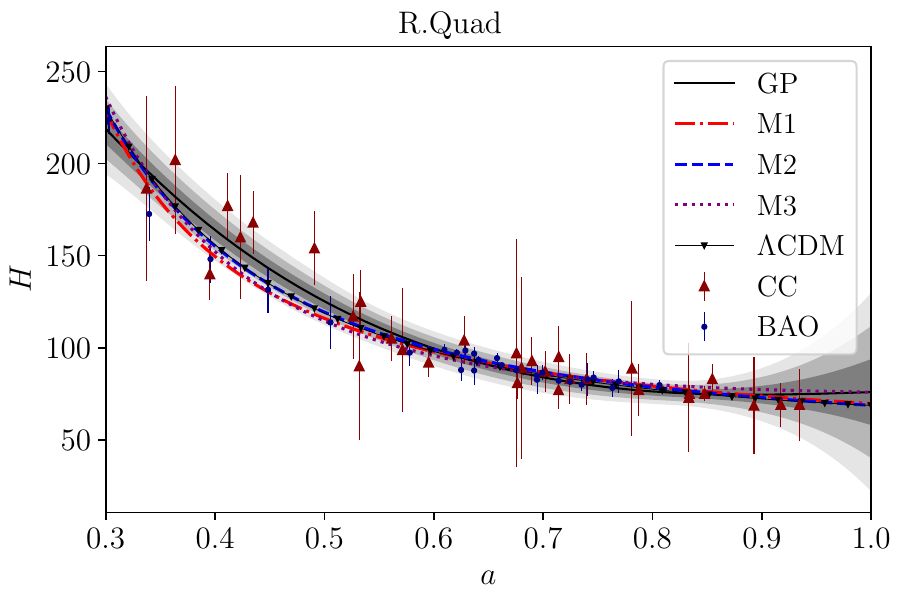}
	\caption{Plots for $H(a)$, in units of km Mpc$^{-1}$ s$^{-1}$, with their $1\sigma$, $2\sigma$ and $3\sigma$ confidence intervals vs $a$ for different choices of the GP kernel,
		where M1, M2 and M3 denote the first, second and third $H(a)$ parametrizations, respectively. }
	\label{fig_gp_H}
\end{figure*}


\noindent As is apparent, a handful of $H(z)$ data are afflicted by
a rather large $1\sigma$ confidence intervals whence some statistically
founded smoothing process must be applied to the full data set to
arrive to sensible results.  This is why we resort to the
machine-learning model Gaussian Process (GP) which infers a
function from labelled training data
\citep{rasmussen2006,seikel2012}, and reconstruct the Hubble
diagram as a function of the scale factor. The said process is
able to reproduce an ample range of behaviors  with just a few
parameters and allows a Bayesian interpretation \citep{zhao2008}.
Then, after smoothing the Hubble data set, we constrain the free
parameters that enter the proposed parametric expressions (Eqs.
(\ref{eq:H(a)}), (\ref{eq:H(a)again}) and
(\ref{eq:H(a)once}), below) of the history of the Hubble
factor  and check whether the corresponding curves, $H(a)$, are
consistent with the second law. That is to say, whether they
comply with the inequality $\, 1+q \geq \Omega_{k}$. \\

\noindent To undertake the GP, we consider the squared
exponential (SqExp hereafter), Mat\'{e}rn 7/2 (Mat72 hereafter),
Cauchy and rational quadratic (R.Quad hereafter) covariance
functions,
\begin{widetext}
	\begin{eqnarray}
	k(a, \tilde{a}) = \begin{cases}
	~~\sigma_f^2 \exp \left( - \frac{(a-\tilde{a})^2}{2l^2}\right) & \text{$\cdots$ Sq.Exp}, \\
	~~\sigma_f^2 \exp \left( - \frac{7 \vert a - \tilde{a} \vert}{l} \right) \left[ 1 + \frac{7 \vert a - \tilde{a} \vert}{l} + \frac{14 \left( a - \tilde{a} \right)^2}{5l^2} + \frac{7\sqrt{7} \vert a - \tilde{a} \vert ^3}{15l^3} \right] & \text{$\cdots$ Mat.72}, \\
	~~\sigma_f^2 \left[ \frac{l}{(a-\tilde{a})^2 + l^2}\right] & \text{$\cdots$ Cauchy}, \\
	~~\sigma_f^2 \left[ 1 + \frac{(a-\tilde{a})^2}{2 \alpha l^2} \right]^{-\alpha} & \text{$\cdots$ R.Quad} ,
	\end{cases}
	\label{eq:covariances}
	\end{eqnarray}
\end{widetext}
\noindent (also called ``kernels") where $\sigma_f$, $l$ and $\alpha$ are the hyperparameters. Throughout this work, we assume a zero mean
function to characterize the GP. We investigate if the different covariance functions lead to significant differences
in the results\footnote{For details on GP, visit \url{http://www.gaussianprocess.org}}. \\

\noindent We adopt a python implementation of the ensemble sampler
for MCMC, the
\texttt{emcee}\footnote{\url{https://github.com/dfm/emcee}},
introduced by \citet{emcee}. The two
dimensional confidence contours showing the uncertainties along
with the one dimensional marginalized posterior probability
distributions, are shown in Fig. \ref{hyper_plot} with the help of the
GetDist\footnote{\url{https://github.com/cmbant/getdist}} module
of python, developed by \citet{getdist}.


\begin{figure*}
	\centering
	\includegraphics[width=0.485\textwidth]{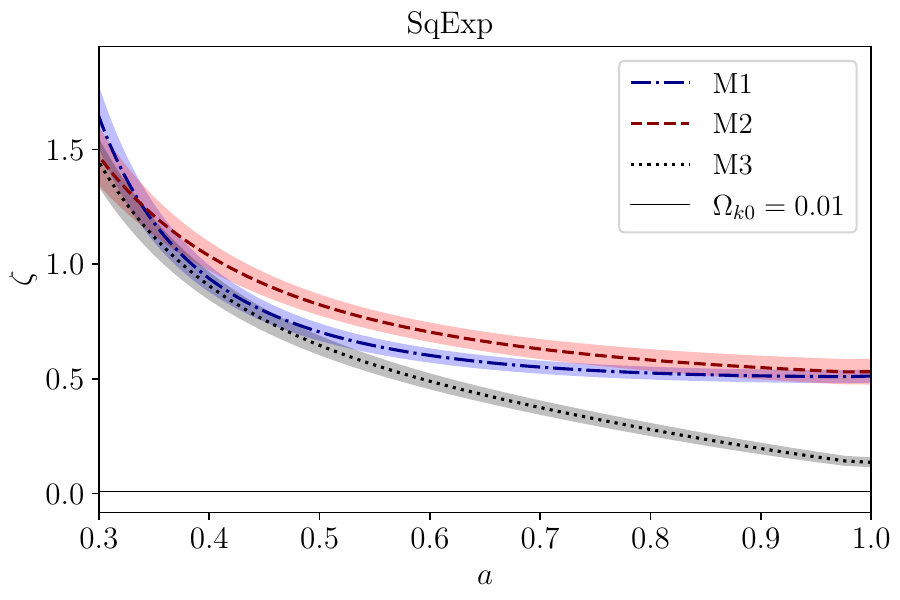}
	\includegraphics[width=0.485\textwidth]{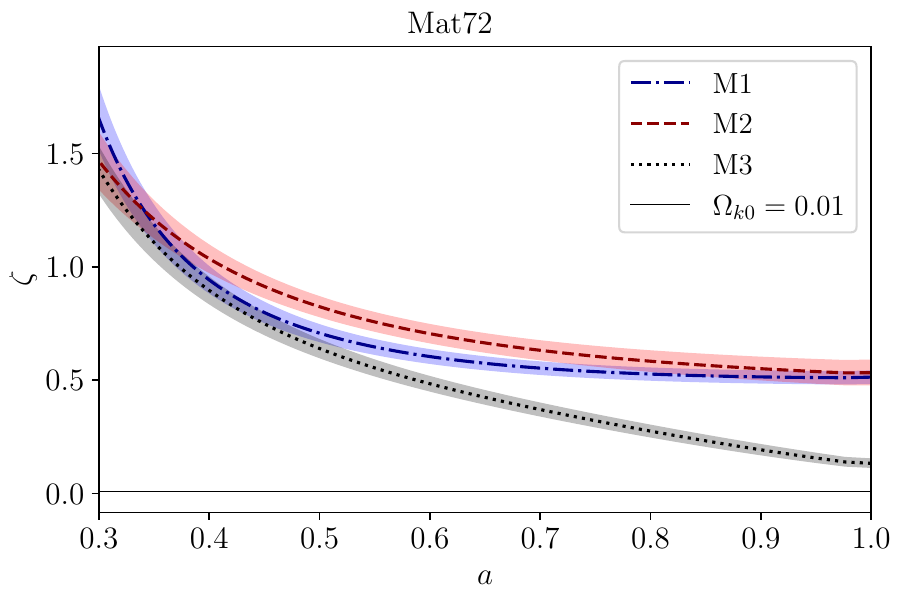}\\
	\includegraphics[width=0.485\textwidth]{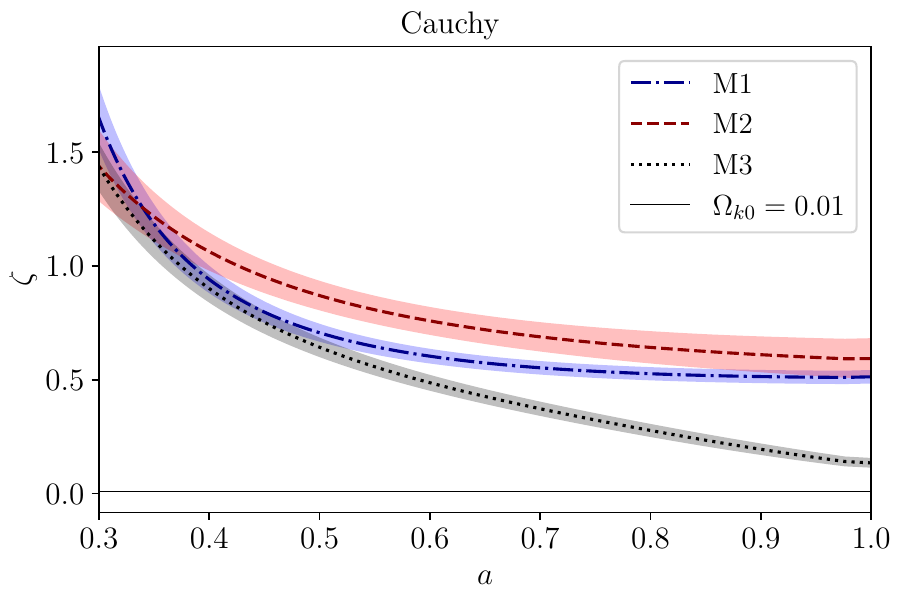}
	\includegraphics[width=0.485\textwidth]{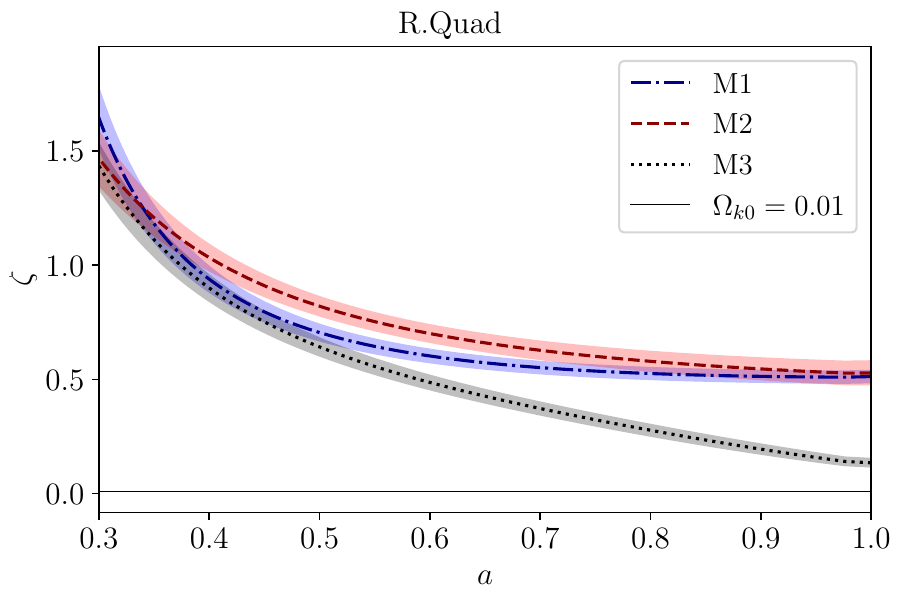}
	\caption{Plots for $\zeta$  with their $1\sigma$ confidence interval vs $a$  for different choices of the GP kernel,
		where M1, M2 and M3 denotes the first, second and third $H(a)$ parametrizations,
		respectively. Every single curve stays above of the observational upper bound of $\Omega_{k0}$, $\sim 10^{-2}$, in the whole scale factor interval.}
	\label{fig_condition}
\end{figure*}


\section{Parametrizations} \label{sec:4}
\noindent In this section we study whether hyperbolic, i.e., open,
spatial sections (which correspond to $ k = -1$) are compatible
with the second law of thermodynamics  as expressed by Eq,
(\ref{q_inequality}). To this end we consider three
parametrizations (equations (\ref{eq:H(a)}), (\ref{eq:H(a)again})
and (\ref{eq:H(a)once}), below) of the history of the Hubble
factor and use the corresponding parameter values obtained after
the smoothing of the Hubble data (tables \ref{tab:M1_table},
\ref{tab:M2_table} and \ref{tab:M3_table} for the first, second
and third parametrizations, respectively\footnote{In these tables
$a_{t}$ is the scale factor at which the transition from
deceleration to acceleration occurs.}). Figure \ref{fig:contour} shows 
the contour plots of the different parameters occurring the three 
parametrizations. Then we check whether the corresponding curves, 
$H(a)$, comply with the inequality (\ref{q_inequality}). These curves 
are shown in Fig. \ref{fig_gp_H}. \\

\noindent We express $\Omega_{k}$ as
\begin{equation}
\Omega_{k} = \Omega_{k0} \left(\frac{a_{0} \,H_{0}}{a \,H}
\right)^{2}, \label{eq:omegakk}
\end{equation}
with $\Omega_{k0} \lesssim 10^{-2}$ the upper bound on the likely
present value of the spatial curvature parameter assuming $k =-1$,
see
\citet{komatsu2011,planck2013,ratra2018,handley2021,narayan2022,bel2022}. \\

\noindent In the cases considered below $H_{\star} \equiv H(a
\rightarrow \infty)$. The analysis of the fourth parametrization does not call for a
numerical study. \\

\noindent Before going any further, it is worthwhile
to note that since $0 < \Omega_{k = -1} < 1$ and that $q > 0$
($<0$) when the universe is decelerating (accelerating), the said
inequality cannot be violated when the universe is decelerating.


\subsection{First parametrization}
\noindent A reasonable and simple parametrization of the Hubble factor in an
ever-expanding FLRW universe, regardless of the spatial curvature (M1 hereafter), is
\begin{equation}
H = H_{\star} \exp{(\lambda/a)},
\label{eq:H(a)}
\end{equation}
where  $H_{\star}$ and   $\lambda$  are free parameters to be fitted  to the data
by means of the GP method. \\

\noindent Since $1+q = \lambda/a$  there is deceleration when $a <
\lambda$ and acceleration when $a > \lambda$. On the other hand,
recalling that $\Omega_{k}$ is given by (\ref{eq:omegakk}), and after setting $a_{0} = 1$, Eq.
(\ref{q_inequality}) boils down to
\begin{equation}
\frac{\lambda}{a} \geq \Omega_{k0} \, \frac{\exp{[2 \lambda( \frac{a-1}{a}})]}{a^{2}} \, ,
\label{eq:Han1a}
\end{equation}
which can be rewritten as
\begin{equation}
\zeta_ {\text{ M1}} =  \lambda \, a \exp\left[ 2\lambda \left(\frac{1}{a} - 1\right)\right] \geq \Omega_{k0}.
\label{eq:zetam1}
\end{equation}

\noindent Using the best fit values of $H_{\star}$ and $\lambda$, shown  in Table \ref{tab:M1_table} corresponding to each kernel,
we find that $\zeta_{\text{M1}}$ stays above $\Omega_{k0}$ in the full range,
$\, (0.3, 1)$, of the scale factor covered by
the data as shown in Fig. \ref{fig_condition}. Thus Eq. (\ref{eq:zetam1}) is satisfied, i.e., the curve $H(a)$
of the first parametrization respects, by a comfortable margin, the second law of thermodynamics also if $k = -1$. 

\subsection{Second parametrization}
\noindent A somewhat less simple
parametrization (M2 hereafter) is
\begin{equation}
H = H_{\star} (1+ \lambda a^{-n}) \, .
\label{eq:H(a)again}
\end{equation}

This one contains three free parameters, namely, $H_{\star}$, $\lambda$ and $n$,
to be fitted to the data. Proceeding as before we can write,

\begin{equation}
\zeta_{\text{ M2}} = \lambda \frac{\left(\lambda +
	a^{n}\right)}{\left[\left(1+\lambda\right)a^{n-1}\right]^2} \geq \Omega_{k0}.
\label{eq:zetam2}
\end{equation}

\noindent Using the best fit values of $\lambda$ and $n$ displayed in table \ref{tab:M2_table}
we see, as  Fig. \ref{fig_condition} shows, that last equation is fulfilled (and therefore
Eq. (\ref{q_inequality})) by an ample margin in the  whole interval of the scale factor
currently observationally accessible. Again, the second  law is respected even if $k=-1$.

\subsection{Third parametrization}
\noindent Similarly, consider the expression (M3 hereafter) for the Hubble rate
\begin{equation}
H(a) = \frac{H_{\star}}{1 - \exp(- \lambda a^{2})}
\label{eq:H(a)once},
\end{equation}
where  $H_{\star}$ and $\lambda$ are free parameter to be fit to the
data.

\noindent Proceeding as in the previous cases from equations
(\ref{q_inequality}) and (\ref{eq:omegakk}) we obtain
\begin{equation}
\zeta_{\text{M3}} = 2 \, \frac{\lambda (1-\exp(-\lambda))^{2}
	a^{4}}{(1-\exp(-\lambda a^{2}))^{3}} \geq \Omega_{k0}.
\label{eq:zetam3}
\end{equation}

\noindent As before, resorting to the best fit values of the free
parameters as given in table \ref{tab:M3_table} it is seen, that the above equation is
satisfied (and consequently Eq. (\ref{q_inequality}) as well) by a
generous margin in the  whole interval of the scale factor
observationally accessible, as shown in Fig. \ref{fig_condition}. Once again, the second  law is
respected even if $k=-1$. 


\begin{figure*}
	\begin{minipage}{0.325\textwidth}
		\begin{minipage}{\textwidth}
			\includegraphics[width=\textwidth]{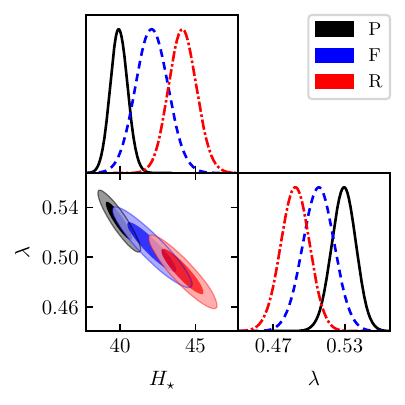}\\ \vspace{-0.8cm}
			\begin{center}
				(a) M1
			\end{center}
			\label{fig:contour_M1_H0}
		\end{minipage}
		\begin{minipage}{\textwidth}
			\includegraphics[width=\textwidth]{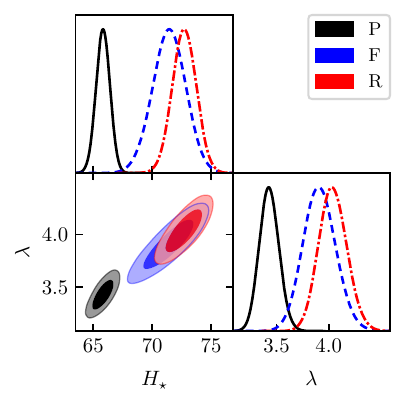}\\ \vspace{-0.8cm}
			\begin{center}
				(c) M3
			\end{center}
			\label{fig:contour_M3_H0}
		\end{minipage}
	\end{minipage}
	\begin{minipage}{0.625\textwidth}
		\includegraphics[width=\textwidth]{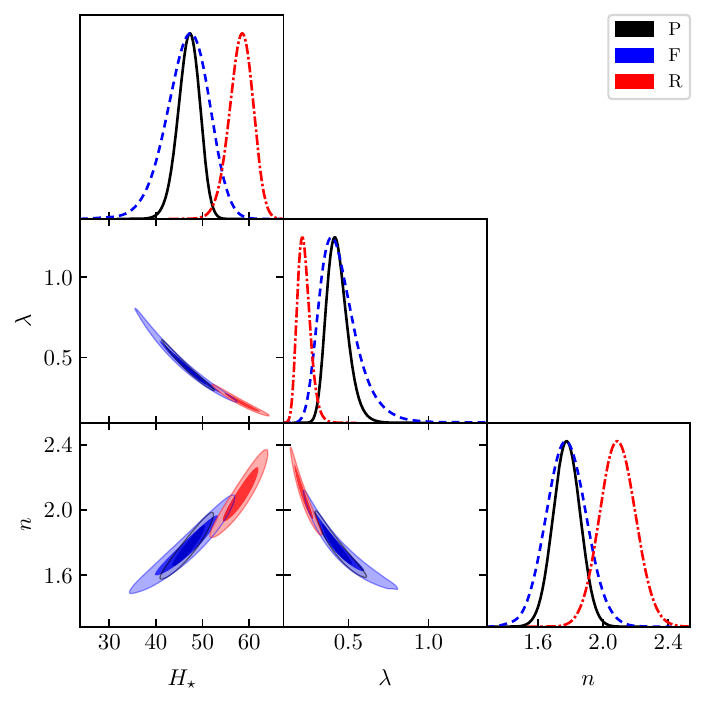}\\ \vspace{-0.8cm}
		\begin{center}
			(b) M2
		\end{center}
		\label{fig:contour_M2_H0}
	\end{minipage}%
	\hfill
	\caption{Contour plots for (a) first $H(a)$ model parameters, (b) second $H(a)$
		model parameters, and (c) third $H(a)$ model parameters, using P20, F21
		and R22 $H_0$ values as priors, where $H_\star$ is in units of km Mpc$^{-1}$ s$^{-1}$.} \label{fig:contour_H0}
\end{figure*}


\begin{table*}
	\caption{{\small Parameter values for the first (M1) parametrization corresponding to the three $H_0$ values as priors, where $H_\star$ and $H_0$ are in units of km Mpc$^{-1}$ s$^{-1}$.}}
	\begin{center}
		\resizebox{\textwidth}{!}{\renewcommand{\arraystretch}{1.15} \setlength{\tabcolsep}{20 pt} \centering
			\begin{tabular}{l c c c c c}
				\hline 
				Prior & $H_{\star}$ & $\lambda$ & $H_0$ & $q_0$ & $a_{t}$\\
				\hline
				P & $39.927^{+0.579}_{-0.566}$ & $0.529^{+0.010}_{-0.010}$ & $67.778^{+0.466}_{-0.458}$ & $-0.468^{+0.010}_{-0.010}$ & $0.529$ \\
				
				F & $42.111^{+1.078}_{-1.058}$ & $0.508^{+0.013}_{-0.013}$ & $69.965^{+1.013}_{-1.001}$ & $-0.489^{+0.012}_{-0.012}$ & $0.508$\\
				
				R & $44.144^{+0.928}_{-0.910}$ & $ 0.488^{+0.012}_{-0.012}$ & $71.937^{+0.797}_{-0.795}$ & $-0.509^{+0.011}_{-0.012}$ & $0.489$ \\
				
				\hline 
			\end{tabular}
		}
	\end{center}
	\label{tab:M1_H0prior}
\end{table*}

\begin{table*}
	\caption{{\small Parameter values for the second (M2) parametrization corresponding to the three $H_0$ values as priors, where $H_\star$ and $H_0$ are in units of km Mpc$^{-1}$ s$^{-1}$.}}
	\begin{center}
		\resizebox{\textwidth}{!}{\renewcommand{\arraystretch}{1.15} \setlength{\tabcolsep}{15 pt} \centering
			\begin{tabular}{l c c c c c c}
				\hline 
				Prior & $H_{\star}$ & $\lambda$ & $n$ & $H_0$ & $q_0$ & $a_{t}$\\
				\hline
				P & $47.105^{+2.216}_{-2.408}$ & $0.426^{+0.070}_{-0.058}$ & $1.776^{+0.083}_{-0.083}$ & $67.194^{+0.477}_{-0.500}$ & $-0.465^{+0.031}_{-0.030}$ & $0.536$ \\
				
				F & $46.948^{+4.256}_{-4.774}$ & $0.428^{+0.133}_{-0.095}$  & $1.774^{+0.123}_{-0.121}$  &  $67.081^{+1.321}_{-1.379}$ & $-0.464^{+0.054}_{-0.051}$ & $0.535$\\
				
				R & $58.330^{+2.417}_{-2.607}$ & $0.219^{+0.043}_{-0.036}$  & $2.090^{+0.111}_{-0.108}$ &  $71.126^{+0.904}_{-0.917}$ & $-0.619^{+0.034}_{-0.033}$ & $0.504$ \\
				
				\hline 
			\end{tabular}
		}
	\end{center}
	\label{tab:M2_H0prior}
\end{table*}

\begin{table*}
	\caption{{\small Parameter values for the third (M3) parametrization corresponding to the three $H_0$ values as priors, where $H_\star$ and $H_0$ are in units of km Mpc$^{-1}$ s$^{-1}$.}}
	\begin{center}
		\resizebox{\textwidth}{!}{\renewcommand{\arraystretch}{1.15} \setlength{\tabcolsep}{20 pt} \centering
			\begin{tabular}{l c c c c c}
				\hline 
				Prior & $H_{\star}$ & $\lambda$ & $H_0$ & $q_0$ & $a_{t}$\\
				\hline
				P & $65.824^{+0.589}_{-0.590}$ & $3.427^{+0.094}_{-0.091}$ & $68.032^{+0.477}_{-0.483}$ & $-0.758^{+0.015}_{-0.016}$ & $0.605$ \\
				
				F & $71.444^{+1.393}_{-1.402}$ & $3.909^{+0.158}_{-0.152}$ & $72.925^{+1.230}_{-1.245}$ & $-0.830^{+0.017}_{-0.019}$ & $0.567$\\
				
				R & $72.722^{+1.014}_{-1.009}$ & $4.037^{+0.136}_{-0.130}$ & $74.025^{+0.914}_{-0.901}$ & $-0.845^{+0.014}_{-0.014}$ & $0.558$ \\
				
				\hline 
			\end{tabular}
		}
	\end{center}
	\label{tab:M3_H0prior}
\end{table*}

\begin{figure*}
	\centering
	\includegraphics[width=0.32\textwidth]{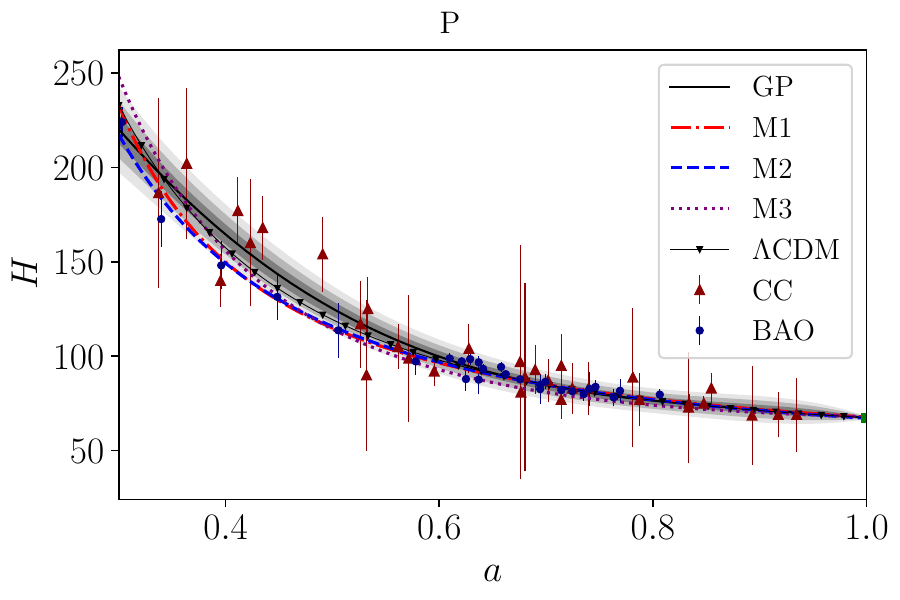}
	\includegraphics[width=0.32\textwidth]{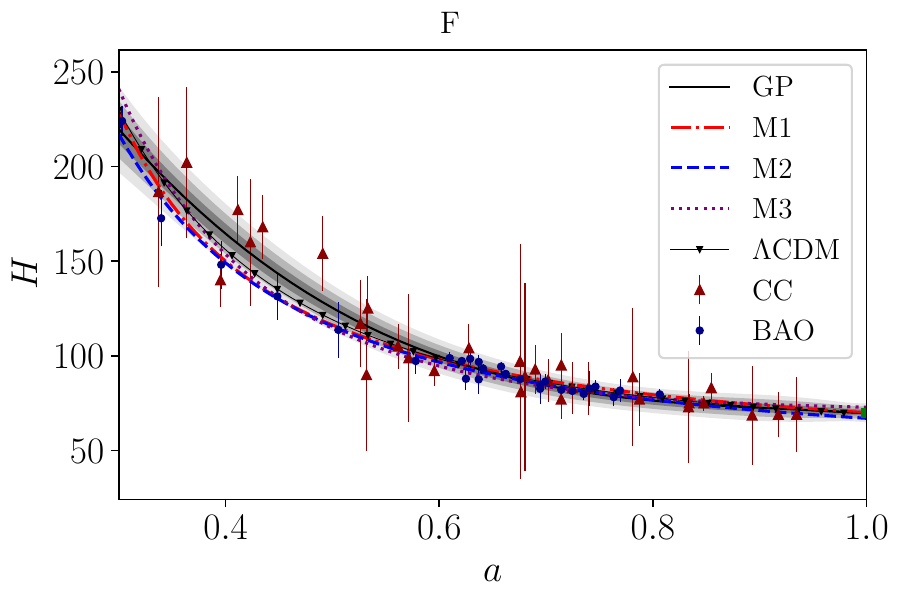}
	\includegraphics[width=0.32\textwidth]{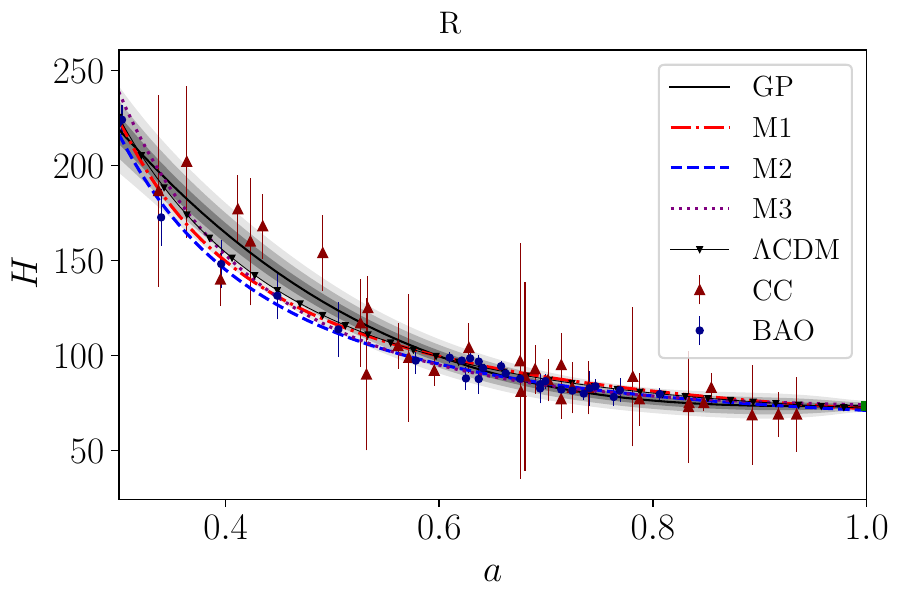}
	\caption{Plots for $H(a)$, in units of km Mpc$^{-1}$ s$^{-1}$, with their $1\sigma$, $2\sigma$ and $3\sigma$ confidence intervals vs $a$ using P20, F21 and R22 $H_0$ values
		as priors (from left to right), where M1, M2 and M3 denotes the first, second and third $H(a)$ parametrizations,
		repectively. }
	\label{fig_gp_H0}
\end{figure*}

\begin{figure*}
	\centering
	\includegraphics[width=0.32\textwidth]{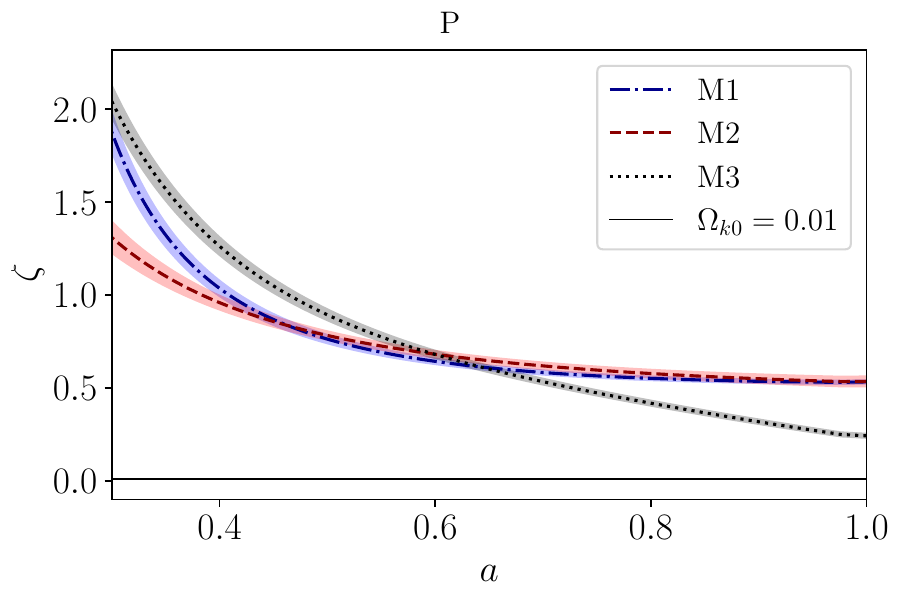}
	\includegraphics[width=0.32\textwidth]{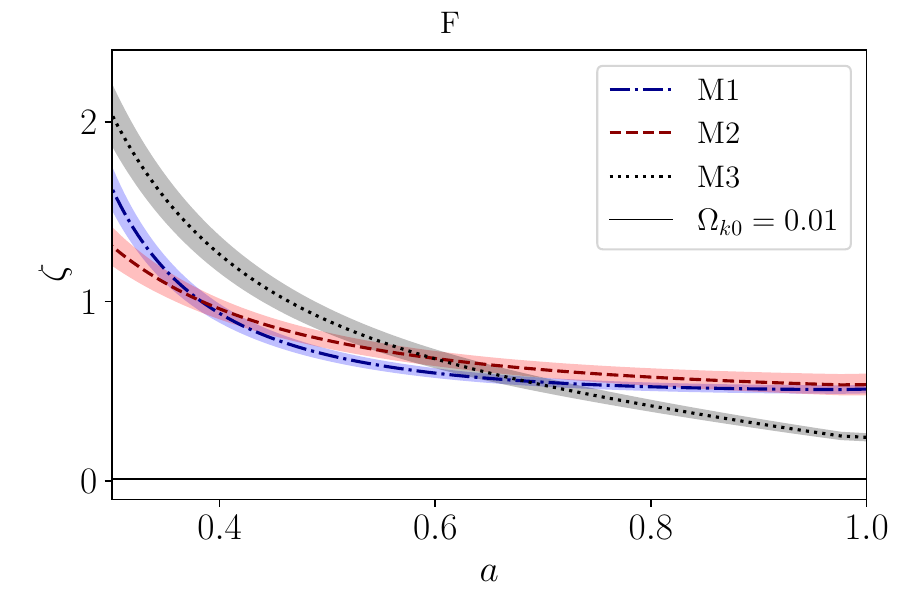}
	\includegraphics[width=0.32\textwidth]{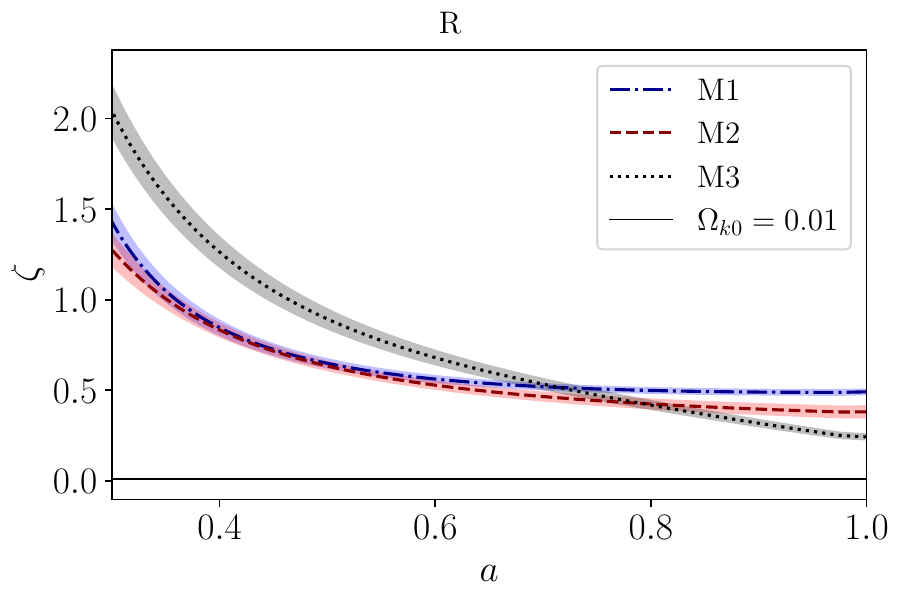}
	\caption{Plots for $\zeta$ with their $1\sigma$ confidence interval vs $a$ using P20, F21 and R22 $H_0$ values as priors
		(from left to right), where M1, M2 and M3 denote the first, second and third $H(a)$ parametrizations, respectively. }
	\label{fig_cond_H0}
\end{figure*}


\subsection{Fourth parametrization}
\noindent Consider the expression for the Hubble factor inspired
in the $\Lambda$CDM model
\begin{equation}
H(a) = H_{0} \sqrt{\Omega_{m0}\, a^{-3} \, + \, \Omega_{k0}\, a^{-2} \, + \, \Omega_{\Lambda}},
\label{eq:H(a)onceagain}
\end{equation}
where $H_{0}$, $\Omega_{m0}$ and $\Omega_{\Lambda}$ are free
parameters. Following parallel steps to the previous cases, we
write $\, \zeta_{\text{ M4}} =  \textstyle{1\over{2}} \, (3
\Omega_{m0}\, a^{-1}\, + \, 2 \Omega_{k0}) \geq
\Omega_{k0}$, that is to say

\begin{equation}
\zeta_{\text{ M4}}= \textstyle{3\over{2}} \Omega_{m0}\, a^{-1}
\geq 0. \label{eq:zetam4}
\end{equation}
It is immediately seen that Eq. (\ref{q_inequality}) is satisfied
for whatever value of the scale factor and consequently the second
law as well. 


\section{Effect of $H_0$ priors} \label{sec:5}
\noindent We further examine to what extent (if at all) the rising
tension between the local measurements of the Hubble constant
\citep{riess2021, freedman2021}, and its inferred values via an
extrapolation of data on the early universe \citep{aghanim2018}
affects our results. \\

\noindent To this end we consider $(i)$ the local measurements by
the SH0ES team [$H_0^{\text{R22}}$ = $73.3 \pm 1.04$ km Mpc$^{-1}$
s$^{-1}$ (R hereafter)], $(ii)$ the updated TRGB calibration from
the Carnegie Supernova Project [$H_0^{\text{F21}}$ = $69.8 \pm
1.7$ km Mpc$^{-1}$ s$^{-1}$ (F hereafter)], and $(iii)$ the
inferred value via an extrapolation of data on the early universe
by the Planck survey [$H_0^{\text{P20}}$ = $67.4 \pm 0.5$ km
Mpc$^{-1}$ s$^{-1}$ (P hereafter)], respectively. We assume
Gaussian prior distributions with the mean and variances
corresponding to the best-fit and $1\sigma$ confidence interval
reported about $H_0$ values. \\

Fig. \ref{fig:contour_H0} shows the contour plots of the different parameters 
occurring in the three parametrizations, on assuming these $H_0$ values. The
constraints on the parameter values are given in Tables \ref{tab:M1_H0prior}, 
\ref{tab:M2_H0prior}, and \ref{tab:M3_H0prior}, for the first, second and third 
parametrizations. The corresponding $H(a)$ and $\zeta$ curves are shown in 
Figs. \ref{fig_gp_H0} and \ref{fig_cond_H0}, respectively. \\

\noindent We can conclude this section by saying that our overall
result, that the constrain $1+q \geq \Omega_{k}$ is satisfied even
if $k = -1$, is not affected whether we take the $H_{0}$ value from
the local measurements or the one obtained from extrapolating the 
precise data drawn from the cosmic microwave background radiation. 

\section{Concluding remarks} \label{sec:6}
\noindent The second law of thermodynamics applied to our
expanding FLRW universe implies that the area of its apparent
horizon cannot decrease, i.e., $ {\cal A}' \geq 0 $. This, in its
turn, dictates $H H' \leq k/a^{3}$. Since in absence of phantom 
fields, which are unstable, $H' < 0$ leads to two possibilities, namely, 
$k = 0$ and $ k = +1$,  result immediately compatible with the second
law. The third possibility, $k = -1$,  may or may not be
compatible with it. Here we resorted to the set of observational
data regarding the evolution of the Hubble factor to discern
whether the said law is satisfied when $k = -1$.  So, with the
help of the non-parametric Gaussian Process we smoothed the set of
$60$ currently available data of the Hubble factor and introduced
four different parametrizations of the ensuing curve, $H(a)$.
After using the latter to constrain the free parameters entering
the first three parametrizations (the fourth one was not in need
of a numerical study), we examined whether the resulting
expressions, with the curvature index $k$ fixed to $-1$, satisfy
Eq. (\ref{q_inequality}). As it turned out, all of them did (Fig.
\ref{fig_condition} and Eq. (\ref{eq:zetam4})). Further, a
quick inspection of the said expressions shows that they also
comply with Eq. (\ref{q_inequality}) in the limit $a \rightarrow \infty$.
Therefore, we are led to conclude that the evolution of FLRW
universes with open spatial sections does not appear to conflict
with  the second law of thermodynamics. Regrettably, a final
verdict cannot be attained at this stage since currently we have
no cosmological model that deserves our unreserved confidence. In
fact, the most reliable model, the so-called ``concordance model",
is afflicted by some drawbacks whose relevance may not be small
\citep{drawbacks, valentino2022}. This is why we did resort to 
parametrize the Hubble factor rather than taking it from any 
given model. \\

\noindent One may argue that this conclusion could be reached
from Eq.(\ref{eq:twoomegas}) straightaway. However this is not
so because that equation only concerns the adiabatic evolution
of the fluid that sources the gravitational field. Further, the 
second law does not enter its derivation whereby it cannot be 
ascertained  whether it is respected or violated.

\section*{Acknowledgments}

PM thanks ISI Kolkata for financial support through Research Associateship.


\section*{Data Availability}

Authors can confirm that all relevant source data are included in the article. The data sets generated during and/or analysed during this study are available from the corresponding author on reasonable request.



\bibliographystyle{mnras}

\bibliography{ms_sc_rev} 



\bsp	
\label{lastpage}
\end{document}